\documentstyle[12pt,epsf,a4]{article}

\begin{document}
\setlength{\baselineskip}{24pt}
\renewcommand{\thefootnote}{\fnsymbol{footnote}}

\title{Artificial market model based on deterministic
  agents and derivation of limit of GARCH type process} 
\author{Aki-Hiro Sato$^{a,}$\protect\footnotemark[1] and Hideki
  Takayasu$^{b}$
\\
{$^{a}$\it \small Department of Applied Mathematics and Physics,} \\
{\it \small Graduate School of Informatics, Kyoto University, Kyoto
  606-8501, Japan} \\ 
{$^{b}$\it \small Sony Computer Science Laboratories,} \\
{\it \small 3-14-13 Higashi-Gotanda, Shinagawa-ku, Tokyo 141-0022, Japan} \\ 
}

\date{\empty}

\maketitle
\footnotetext[1]{Corresponding author. Fax +81-75-753-4919. E-mail
  address: aki@amp.i.kyoto-u.ac.jp (A.-H. Sato).} 

\begin{center}
{\it abstract}
\end{center}
\noindent
We investigate the dealer model -- an artificial market model based on
deterministic agents both numerically and theoretically. The agents
refer to the past market prices and changes their ask/bid price. 
The temporal development of the market price fluctuations is
calculated numerically. A probability density function of the market
price changes has power law tails. Autocorrelation 
coefficient of the changes has an anti-correlation, and
autocorrelation coefficient of squared changes (volatility correlation
function) has a long time correlation. A probability density function
of intervals between two successive transactions follows a geometric
distribution. The GARCH type stochastic process is approximately
derived from the market changes of the model in a limit case. We
discuss two factors of the market price fluctuations and display  
a relation between the volatility of the market prices and a
demand-supply curve. We conclude that the power law tails and
the long time volatility result from both positive and negative
heterogenous feedbacks in the agents. 

\noindent

\noindent
{\bf keywords}: artificial market model, GARCH type stochastic
process, power-law distribution, volatility clustering, econophysics \\ 

\section{Introduction}
\label{sec:introduction}
Economically motivated problems are attracting the attention of
physicists, economists and engineers. Recently physicists have become
interested in these problems and established a new field,
``econophysics'', which has been made tremendous progress since
1997. The price fluctuation is one of the most exciting interests in
econophysic~\cite{Stanley:00,Ausloos:00,Gabaix:03a,Gabaix:03b}.
Statistical properties of market price fluctuations were clarified
from analyzing financial time series empirically. Specifically we
focus on two statistical properties of price fluctuations; (1) the
probability distribution (pdf) of price changes has fat
tails~\cite{Gabix:03a}, and (2) the squared or absolute price 
changes have long time correlation, which is well-known as the volatility
clustering~\cite{Pasquini:99}. The volatility clustering is measured
by autocorrelation function of squared or absolute price changes, the
volatility autocorrelation function, which has a long tail when the
volatility is clustered. 

On one hand stochastic models have been proposed in order to describe
these phenomena. For example, the truncated L\'evy
flights~\cite{Mantegna:99,Nakao:00}, the random multiplicative
processes~\cite{Takayasu:97,Sornette:98,Aki-a:00} and the GARCH
models~\cite{Engle:82,Bollerslev:92}. On the other hand, various
agent-based models have been proposed in order to explain the basic
properties of price fluctuations from the viewpoint of complex
dynamical systems. These studies are called a agent-based
approach~\cite{Bak:97,Johnson:98,Lux:99,Challet:01,Jefferies:01}. 

The first agent-based model of a market was introduced by one of the
authors (H.T.) {\it et. al.} in order to explain why market price
apparently fluctuate randomly~\cite{Takayasu:92}. It was shown that
trading involve a kind of likely nonlinear interaction among agents in
general. A pseudo-random walk fluctuation of market price results from 
the effect of chaos in an artificial market. That is verified in the market 
that includes only 3 agents with very simple deterministic
strategies. Also, it was pointed out that when the agents have the
tendency of following the latest market trend, large fluctuations such
as crashes or bubbles occur spontaneously. The market model is a
variant of the deterministic agent-based model. 

Bak {\it et. al.} investigated a stock market model with ''rational
traders'' and ``noise traders''. Noise traders' actions depend on
their current volatility in the market and imitate ways to buy or
sell. Rational traders optimize their own utility functions. They
emphasize that it is important for noise traders to exist in order to
have fat tails of the pdf~\cite{Bak:97}. Lux {\it et. al.}
investigated a stochastic multi-agent model with the pool of traders
divided into fundamentalists and noise traders~\cite{Lux:99}. They
reported that a pdf of returns in their model has fat tails. These
models assume that a market has a balance of demand and supply at a
unit time to decide the next market price, and the market price
changes depending on an unbalance of the demand and supply. Unlike the
stochastic model of Lux stimulative extensions to the basic
Minority Game (MG), multi-agent market models without noise trader,
have been developed and are successively
investigated~\cite{Challet:01,Jefferies:01}. In fact these studies
demonstrate the stylized facts (the fat tails of the pdf of market
price changes and the volatility clustering) and the numerical results
show that the model contains the GARCH properties. However
the GARCH process was analytically approximated in these studies.

Specifically GARCH model are useful tools in econometrics. However
there are few studies that give the theoretical basis of the GARCH
from microscopic viewpoint. The aim of the present article is that we
develop an artificial model characterized by a few parameters and
directly derive the GARCH type stochastic process from a price change
of the model. In our preceding study we proposed a dynamical model of a
market having mean field interactions of the dealers~\cite{Aki:97}. We
derived a Langevin equation with both positive and negative feedbacks
in a stochastic manner for market prices and, the pdf of time interval
between successive trading follows exponential decay. But
autocorrelation function of the squared change does not have a long
time correlation. In the present article we introduce heterogeneity of
the dealers instead of the mean-field interaction in our preceding study.

The present article is organized as follows. In Sec. \ref{sec:model}
we introduce an artificial model of a market based on 
deterministic agents who modify their ask/bid price depending on past
price changes. In Sec. \ref{sec:simulation} the statistical properties
of artificial market price fluctuations from numerical simulations are
shown. In Sec. \ref{sec:ARCH-limit} we show how the GARCH type stochastic
process for the market price changes is theoretically derived from the
market model in a limit manner. The GARCH type stochastic process is a
familiar model for time series analysts but its theoretical basis in
microscopic viewpoint has not been clarified yet. In
Sec. \ref{sec:discussion} we discuss the volatility clustering
demonstrated from numerical simulation in the model comparing with the
demand and supply curves of the market. Sec. \ref{sec:conclusion} is
devoted to concluding remarks.

\section{Market model}
\label{sec:model}
We show a brief explanation of the dealer model -- an artificial
market model with many simple deterministic
dealers~\cite{Aki-b:02}. The fundamental idea of the model is an
interaction between agents (dealers) through a common board and 
historical market prices. We consider a market in which $N$ dealers
exchange a single commodity, for example, in foreign currency
market. We assume that the dealers have limited information due to a
short time trading. It takes a finite time for them to decide to buy
or sell. Therefore they have each simple strategy to predict the next
market price from the limited information. By the same token we do not
assume that they have utility function since they cannot optimize
their action in short time. Dealers put a sell/buy order on the
market. By a competitive mechanism in the market adequate orders are
selected, and the next market price is determined from the ordered price.

\begin{figure}[htb]
\centering
\epsfxsize=320pt
\epsfbox{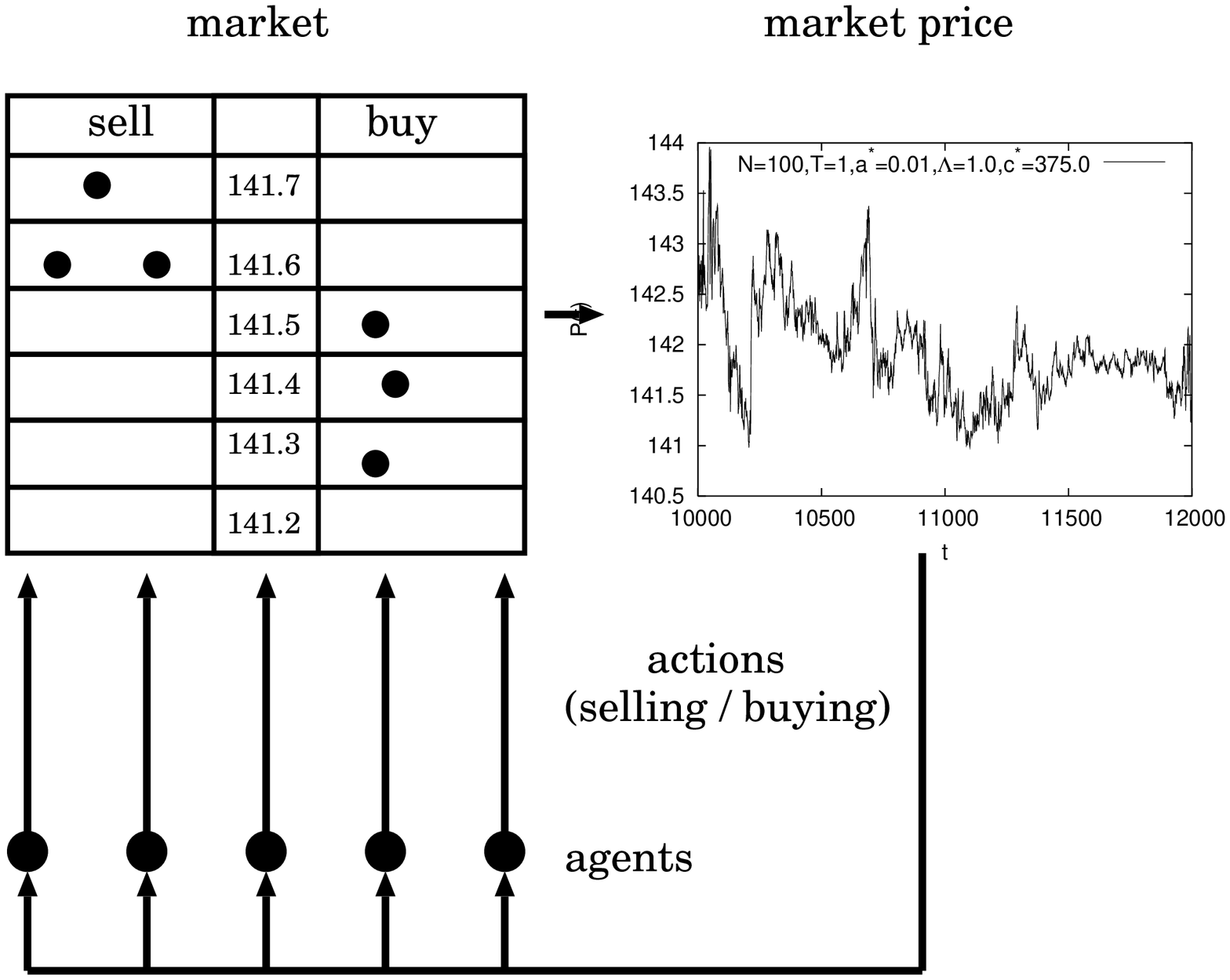}
\caption{The conceptual illustration of the dealer model. The inputs
  of the market are orders from dealers. The output of the market is a
  market price. The input of an agent is a sequence of past price
  changes. The output of an agent is a sell price or a buy price.}
\label{fig:concept}
\end{figure}

As shown in fig. \ref{fig:concept} we explain the dealer model for two
parts: (1) a market rule, which describes how to determine a 
market price from orders, (2) a strategy of the agent, which governs
how to order a sell/buy into the market and how to modify their
ask/bid prices. In the following subsections we explain a market
rule, a strategy of the agent, initial conditions and parameters. 

\subsection{Market rule}
A market has a competitive mechanism for orders. We assume that 
$N$ dealers put their order (sell/buy) to a common board. Here $p_i(t)$
represents an ask/bid price of the $i$th dealer at time $t$. The ask
prices and the bid prices individually compete in the market. Namely 
the maximum buying price and the minimum selling price are effective
in the market. Thus the condition for a trading to occur is given by the
following inequality,
\begin{equation}
\max_{\mbox{\tiny{for all buyers}}}\{p(t)\} \geq \min_{\mbox{\tiny{for
      all sellers}}}\{p(t)\}, 
\end{equation}
where the right hand side represents the maximum bid price for all the
buyers, and the left hand side represents the minimum ask price for
all the sellers. Assume that the market price $P(t)$ is determined as
an arithmetic mean of the buy price and the sell price when trading
occurs. Otherwise the last market price is maintained. Namely,  
\begin{equation}
P(t) = \left\{
\begin{array}{ll}
{\displaystyle \frac{1}{2}\bigl(\mathop{\max\{p(t)\}}_{\mbox{\tiny{for all
          buyers}}} + \mathop{\min\{p(t)\}}_{\mbox{\tiny{for all
        sellers}}} \bigr)} &
{\displaystyle (\mathop{\max\{p(t)\}}_{\mbox{\tiny{for all buyers}}} \geq
  \mathop{\min\{p(t)\}}_{\mbox{\tiny{for all sellers}}})}  
\\ 
P(t-1) &
{\displaystyle (\mathop{\max\{p(t)\}}_{\mbox{\tiny{for all buyers}}} <
\mathop{\min\{p(t)\}}_{\mbox{\tiny{for all sellers}}})}  
\end{array}
\right..
\label{eq:market-mechanism}
\end{equation}

\subsection{Strategy of the dealer}
In general a dealer determines his/her action from several causes. The
causes are separated into ``endogenous'' factors and
``exogenous'' ones. The endogenous ones mean what has happened in
the market and contains the historical market prices and a
rumor in the market. The exogenous ones mean what has happened outside
market, a balance of domestic commerce or international commerce among
countries and so on. Here we consider that the dealer has to make a decision
based on the market price fluctuations since they regards only
prices for a short period and determine their action and ask/bid
price.  

Here we consider a trade between a seller and a buyer. Suppose that the
both must make a trade. Then the seller decreases his/her sell price a
little bit unless he/she can make a trade. On the other hand the buyer
increases his/her buy price a little bit. Repeating this they find a
satisfying exchange price. 

The same is consistent in the market. Sellers go on decreasing their
sell price a little bit until they can sell and buyers go on
increasing their buy price until they can buy. From the assumption a
modification of the ask/bid price is negative/positive for
sellers/buyers. The temporal development of the ask/bid price can be
described by,
\begin{equation}
p_i(t+1) = p_i(t) + \alpha_i(t) D_i(t),
\end{equation}
$\alpha_i(t)$ represents the $i$'th dealer's modification per a unit
time step, and $D_i(t) = -1$ when the $i$th dealer is a seller, and
$D_i(t) = 1$ a buyer. 

Moreover we assume that a value of modification depends on the past
market prices due to limited information. The assumption
means that the next action of dealers is only affected by the
historical data of market prices. 
\begin{equation}
\alpha_i(t) = \alpha(P(t),P(t-1),\ldots)
\end{equation}
We assume that the dealers are sensitive to price changes rather than
exact prices. One of the simplest modification algorithms is a linear
inner product of dealer-dependent coefficients and past price changes.  
Namely the modification $\alpha_i(t)$ is described by 
\begin{equation}
\alpha_i(t) = | 1 + \sum_{s'=1}^{T}c_{i,s'} \Delta P_{prev}(s') | a_i,
\end{equation}
where $\Delta P_{prev}(s')$ denotes the $s'$th change of the past
market price. It is obvious that $\Delta P_{prev}(1)$ is the latest
market price change. $c_{i,s'}$ represents coefficients of the $i$th
dealer for the last $s'$th market price change, $a_i$ is a positive
coefficient.

For simplicity we consider the case of $T=1$. Then a rule to modify
dealers' expectation price is written by 
\begin{equation}
p_i(t+1) = p_i(t) + | 1 + c_i \Delta P_{prev}(1) | a_i  D_i(t),
\label{eq:dealers-rule}
\end{equation}
where $c_i$ is a coefficient, which corresponds to accelerating and
deaccelerating his/her modification of expectation price depending on
the latest price. After the large price change each dealer goes up or
goes down quickly his/her expectation price in order to make his/her
portfolio balanced as soon as possible.

We assume that the dealers open a sell/buy position till they exchange.
They determine whether they open the sell/buy position
after they have made a trade. We assume that the sellers want to go on
selling when the market price goes up and that the buyers want to go on
buying when the market price goes down. 

Therefore after trading the commodity the two dealers who have
exchange open either a sell position or a buy position depending on
the past price change $\Delta P_{prev}$. If the market price goes up
then the dealers may expect to make a profit from selling the
commodity. Hence we assume that they open the buy position, i.e.,
$D_i(t+1)=-1$, when $\Delta P_{prev} > 0$. In striking contrary we
assume that they open the buy position, i.e., $D_i(t+1)=1$ when $\Delta
P_{prev} < 0$ in order to make a profit from future selling the
commodity. Furthermore they open either a sell position or a buy
position at the same probability $p=1/2$ when $\Delta P_{prev} = 0$. 

Moreover we assume that dealers have difference between the market
price and the retried ask/bid price, which denotes $\Lambda_i (\equiv
P(t) - p_i(t))$. In a mind of a dealer it is clear that the retried
ask price is greater than the market price. On the other hand the
retried bid price is less than the market price. Therefore the ask/bid
price of the $i$th dealer at $t$ is written by, 
\begin{equation}
p_i(t) = P(t) - D(t+1)\Lambda_i 
\end{equation}
For simplicity $\Lambda_i = \Lambda$ for the all dealers.

\subsection{Initial conditions and parameters}
Each dealer has two coefficients: $a_i$ and $c_i$. $a_{i}$ and $c_{i}$
are initially given by uniform random numbers distributed in interval $[0
  ,a^*]$ and $[-c^*,c^*]$, respectively. These coefficients exhibit
dealer's personality and are fixed throughout a numerical
simulation. Each dealer has two variables: $D_i(t)$ and
$p_{i}(t)$. Initial condition $D_i(0)$ is given by either $+1$ or $-1$
randomly, {\it i.e.} a probability for $D_i(0)$ to be $1$ is $1/2$,
and a probability for $D_i(0)$ to be $-1$ is $1/2$. Then $p_{i}(0)$ is
given by $P_0-\Lambda$ when $D_i(0)=1$ and $P_0+\Lambda$ when
$D_i(0)=-1$, where $P_0$ is an initial market price fixed as
$P_0=120.0$ throughout simulations. 

We assume that the latest price change $\Delta P_{prev}$ is initially
zero. The dealer's rule is deterministic except initial conditions.  
This model has four parameters; amount of the dealers $N$, $a^*$ for
$a_i$, difference between the ask/bid price and the market price  
$\Lambda$ and $c^*$ for $c_i$. Finally we summarize the model
parameters in Tab. \ref{tab:parameters}. 

\section{Numerical simulation}
\label{sec:simulation}
\begin{figure}[htb]
\centering
\epsfxsize=200pt
\epsfbox{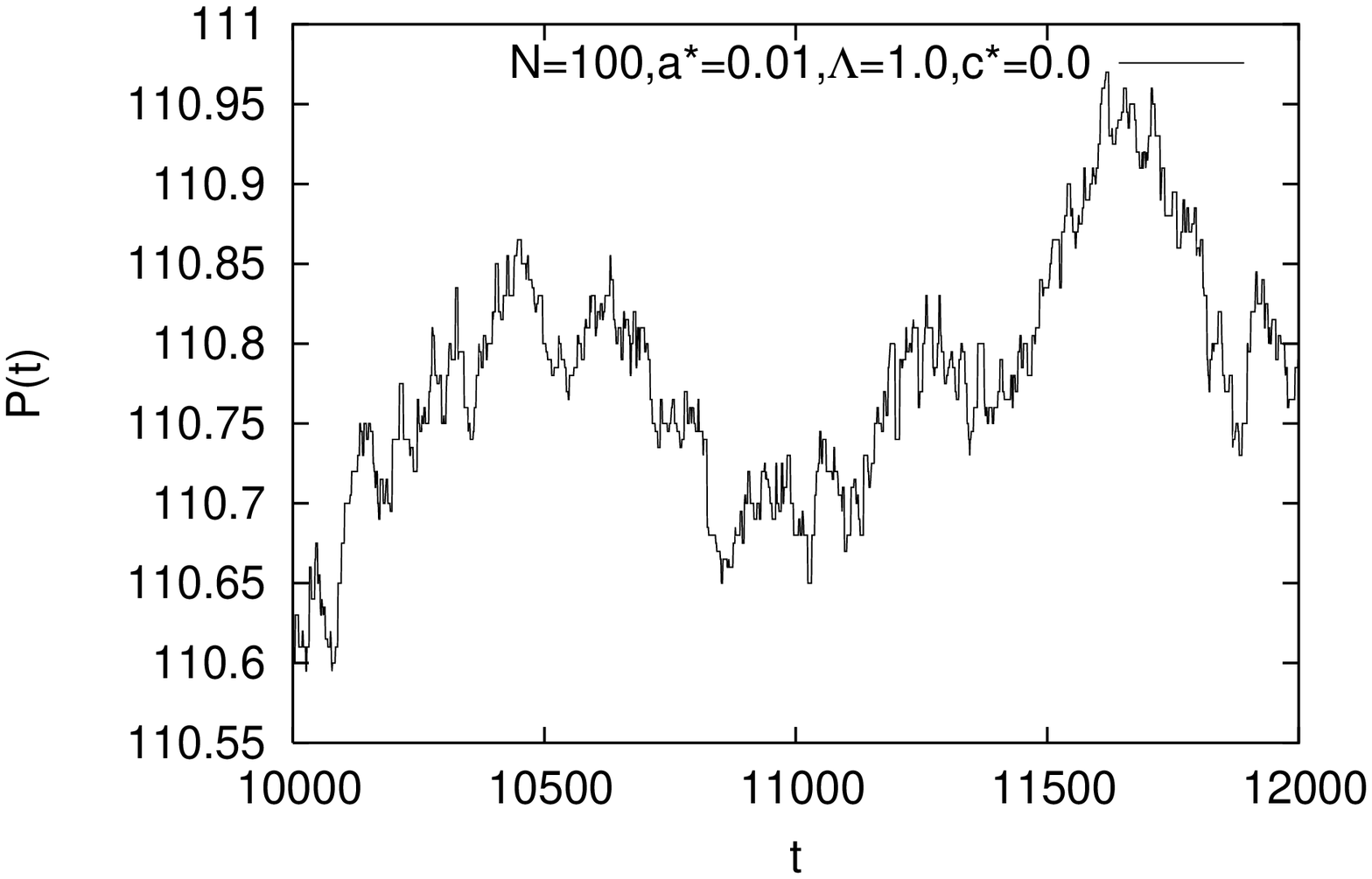}
\epsfxsize=200pt
\epsfbox{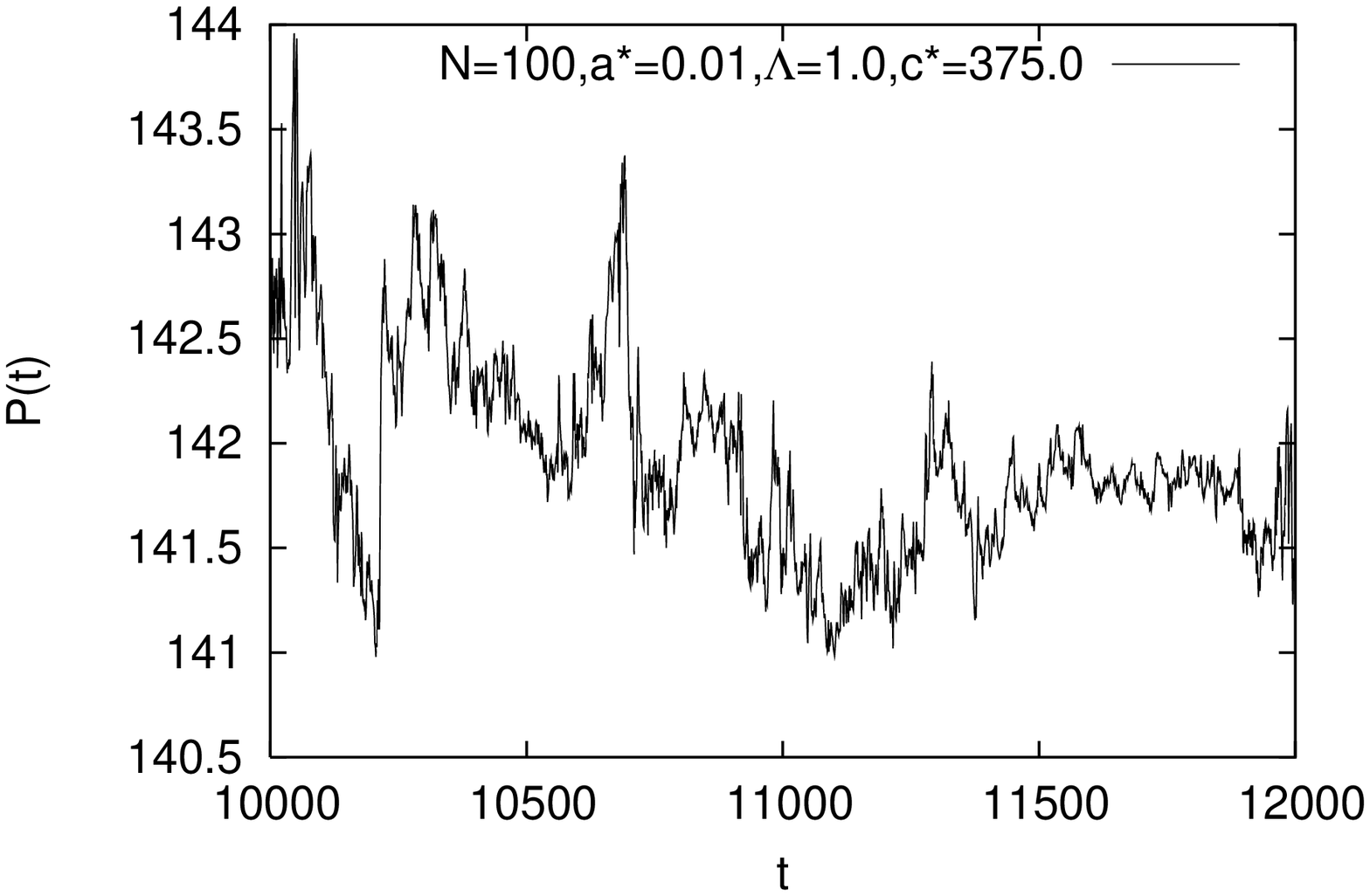}
\caption{A typical example of time series of market prices at $N=100$, $a^*=0.01$ and $\Lambda=1.0$ for $c^*=0.0$(right) and $375.0$(left).} 
\label{fig:market-price}
\end{figure}

\begin{figure}[htb]
\centering
\epsfxsize=200pt
\epsfbox{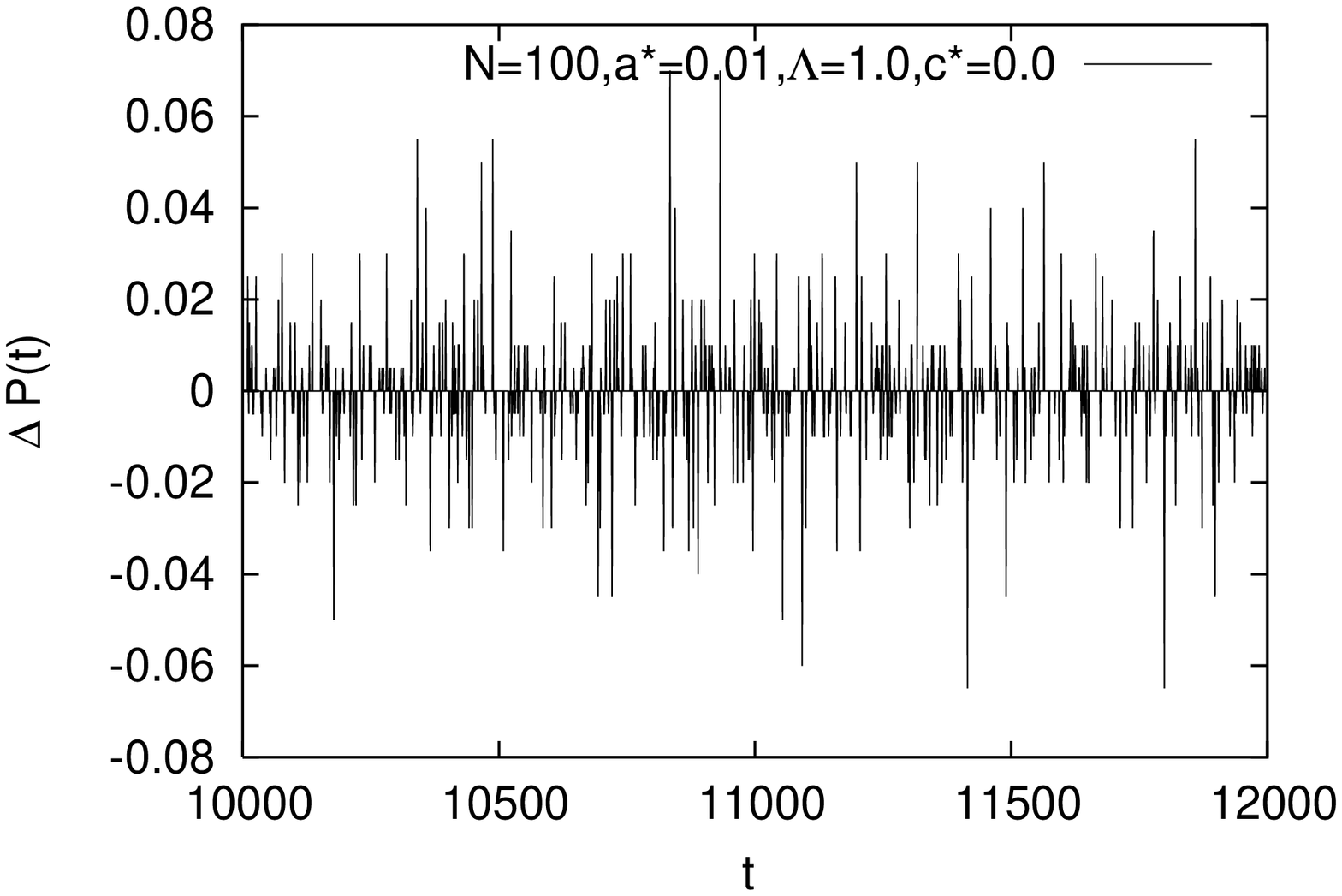}
\epsfxsize=200pt
\epsfbox{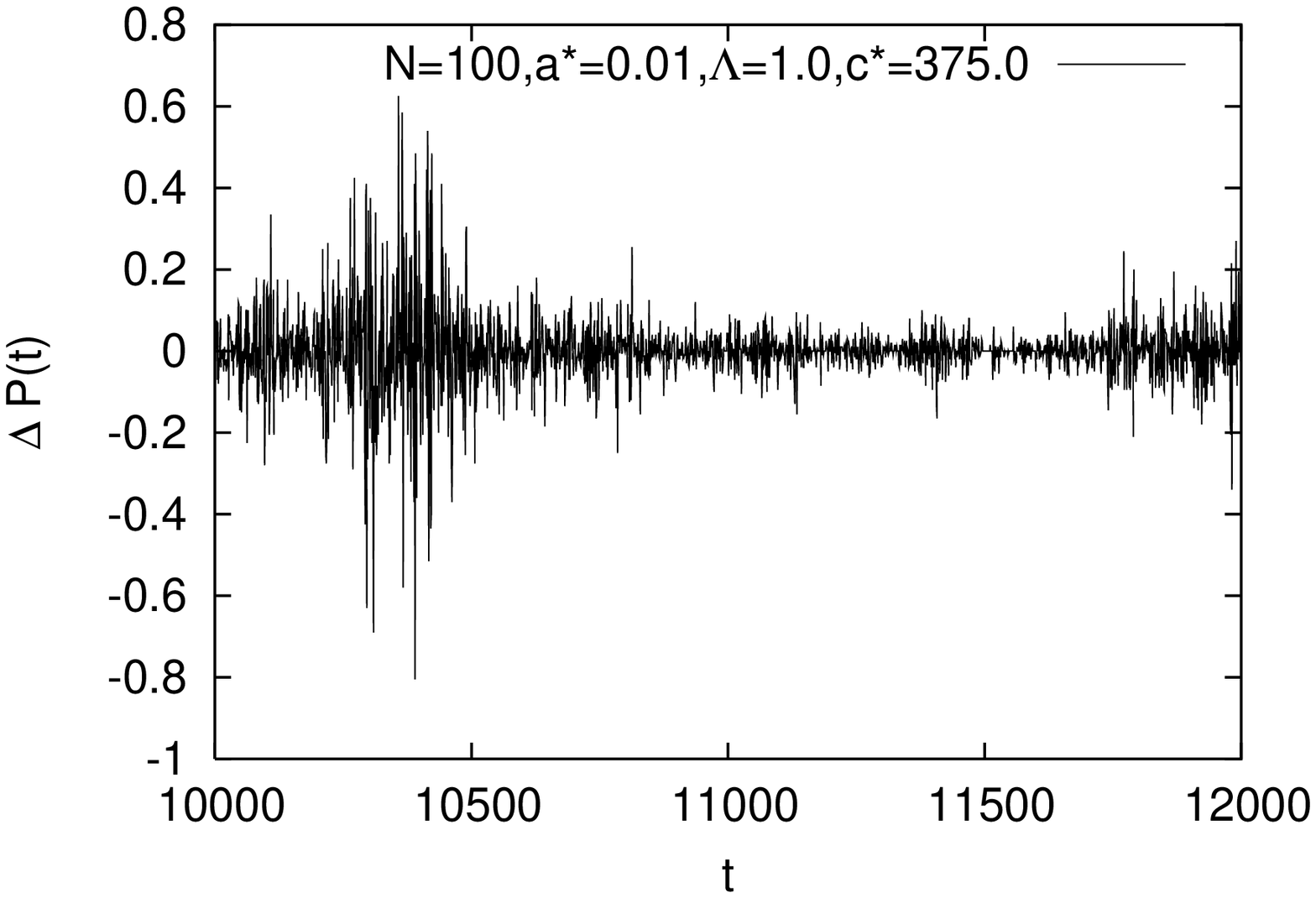}
\caption{A typical example of time series of price changes at $N=100$, $a^*=0.01$ and $\Lambda=1.0$ for $c^*=0.0$(left) and $375.0$(right).} 
\label{fig:market-price-change}
\end{figure}

Fixing parameters $N=100$, $\Lambda=1.0$ and changing $c^*$ and $a^*$,
we numerically simulate the dealer model introduced in the above
section. Figs. \ref{fig:market-price} and
\ref{fig:market-price-change} show a typical example of time series of
market prices $P(t)$ and of price changes $\Delta P(t) = P(t) -
P(t-1)$. The market prices and their changes apparently fluctuate
although the model is completely deterministic. The reason is
high-dimensionality of the system as the freedom of the system is
proportional to the agent-number.  

\begin{figure}[htb]
\centering
\epsfxsize=320pt
\epsfbox{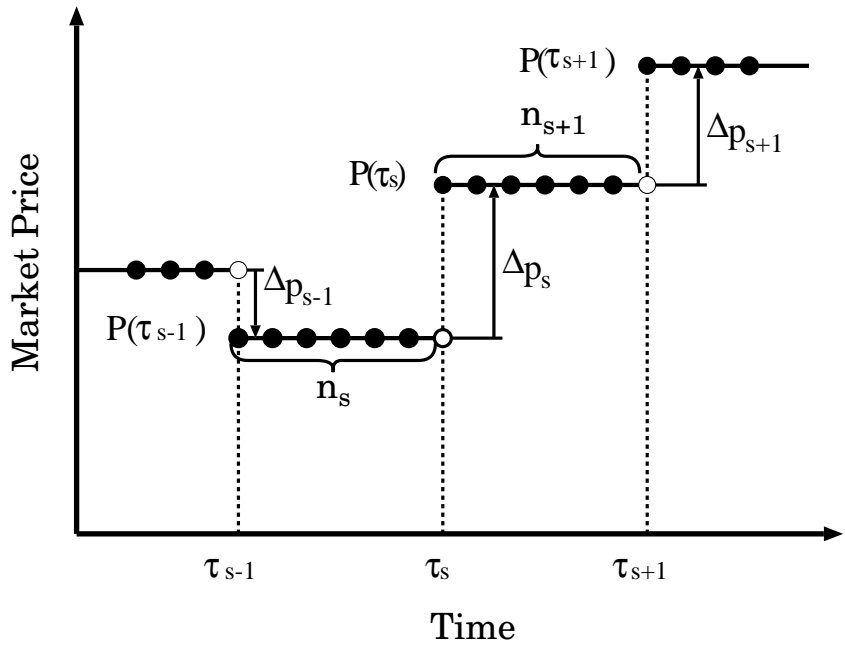}
\caption{A conceptual illustration of a market price fluctuation. A
  trading occurs at random. $\tau_s$ represents the time when the
  $s$th trading occurs. $P(\tau_s)$ exhibits a market price at
  $\tau_s$, $\Delta p_s$ a market price change at $\tau_s$, namely
  $\Delta p_s = P(\tau_s) - P(\tau_{s-1})$, and $n_s$ a difference
  between $\tau_{s-1}$ and $\tau_s$, $n_s \equiv \tau_{s} -
  \tau_{s-1}$.
}
\label{fig:fluctuation}
\end{figure}

Let $\tau_s$ denote the time when the $s$th trade
occurs. Fig. \ref{fig:fluctuation} exhibits a conceptual illustration 
of market price fluctuations. The time series may be characterized by
a price change and a time difference between successive
trading. $P(\tau_s)$ represents a market price at $\tau_s$. $\Delta
p_s$ denotes a price change at $\tau_s$, namely, $\Delta p_s \equiv 
P(\tau_s) - P(\tau_{s-1})$, and $n_s$ a time difference between
successive trading, $n_s \equiv \tau_{s} - \tau_{s-1}$.  

\begin{figure}[htb]
\centering
\epsfxsize=200pt
\epsfbox{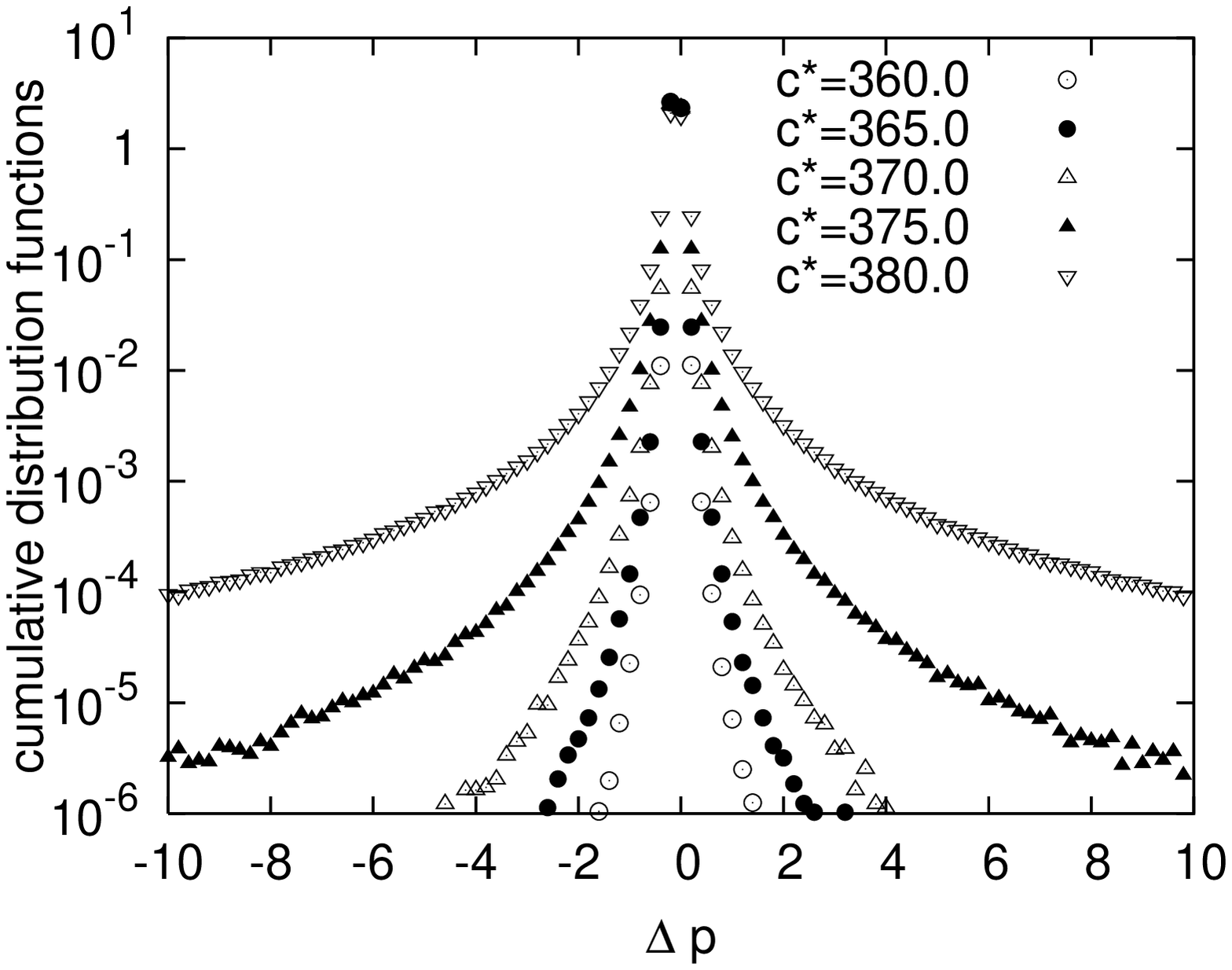}
\epsfxsize=200pt
\epsfbox{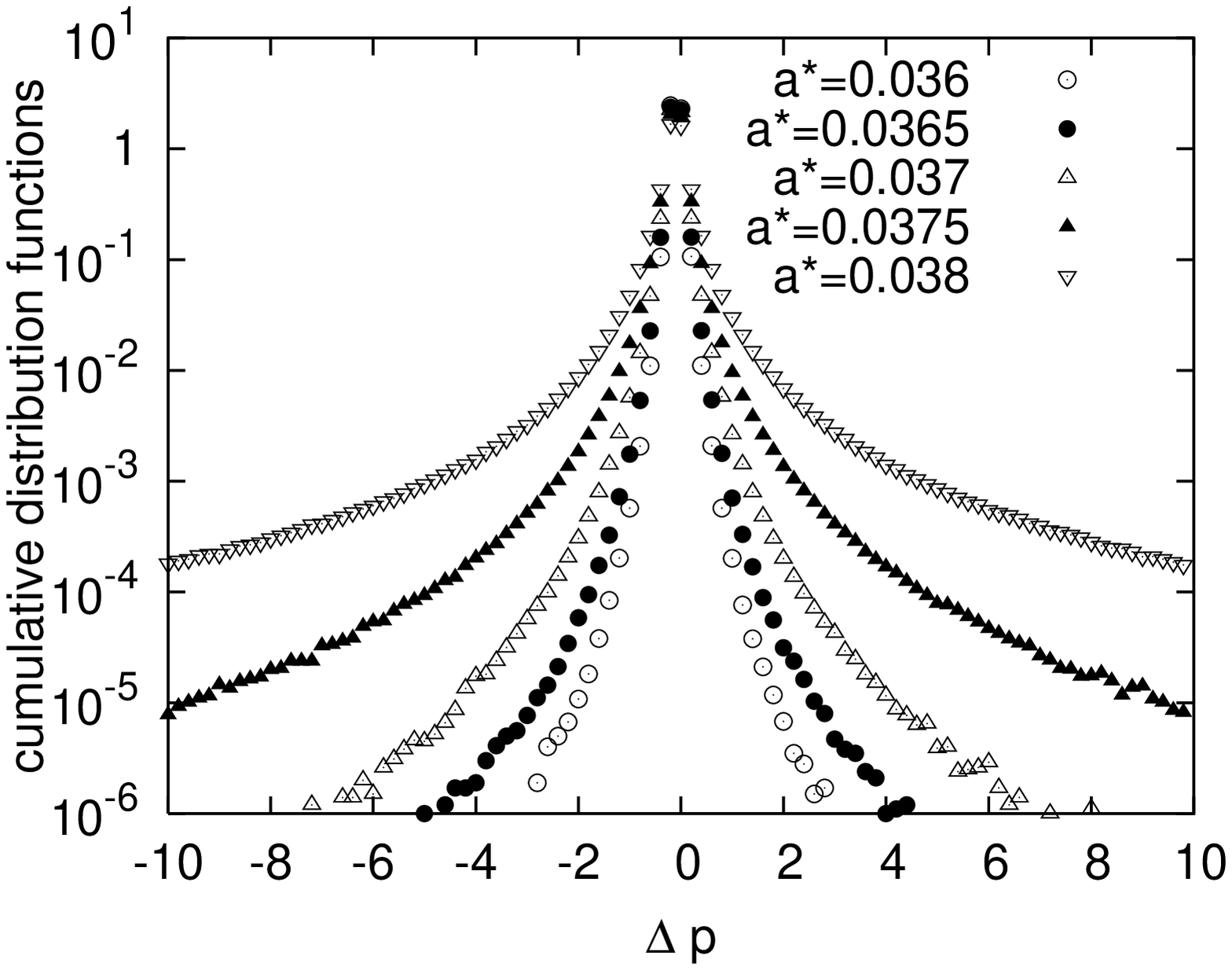}
\caption{Semi-log plots of the probability density functions of the
  price changes $\Delta p_s$. We fix $N=100$, $\Lambda=1.0$ and 
  $a^*=0.01$, and change $a^*$ (left). We fix $N=100$, $\Lambda=1.0$
  and $c^*=100.0$, and change $a^*$ (right). 
}
\label{fig:pdfs}
\end{figure}

\begin{figure}[htb]
\centering
\epsfxsize=200pt
\epsfbox{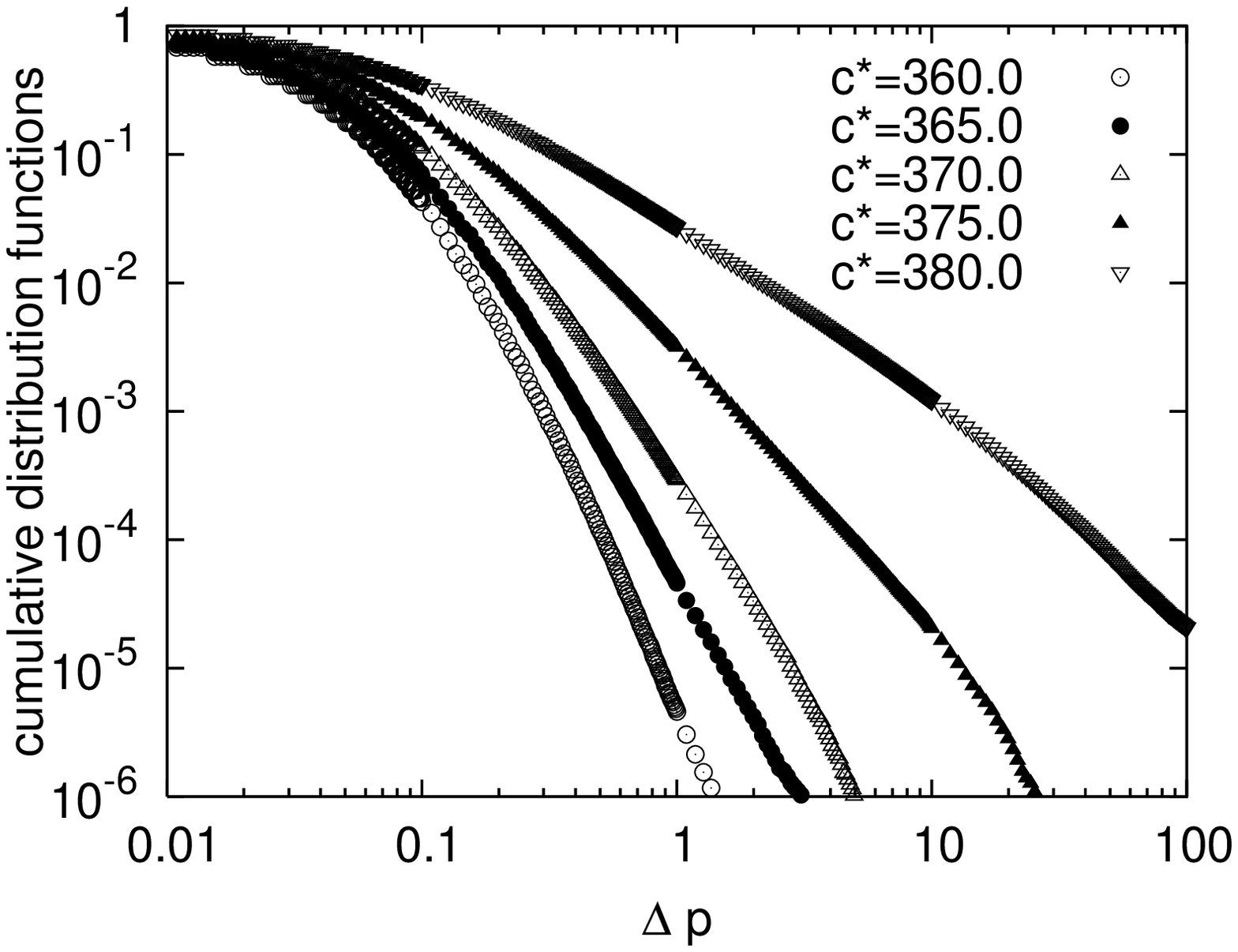}
\epsfxsize=200pt
\epsfbox{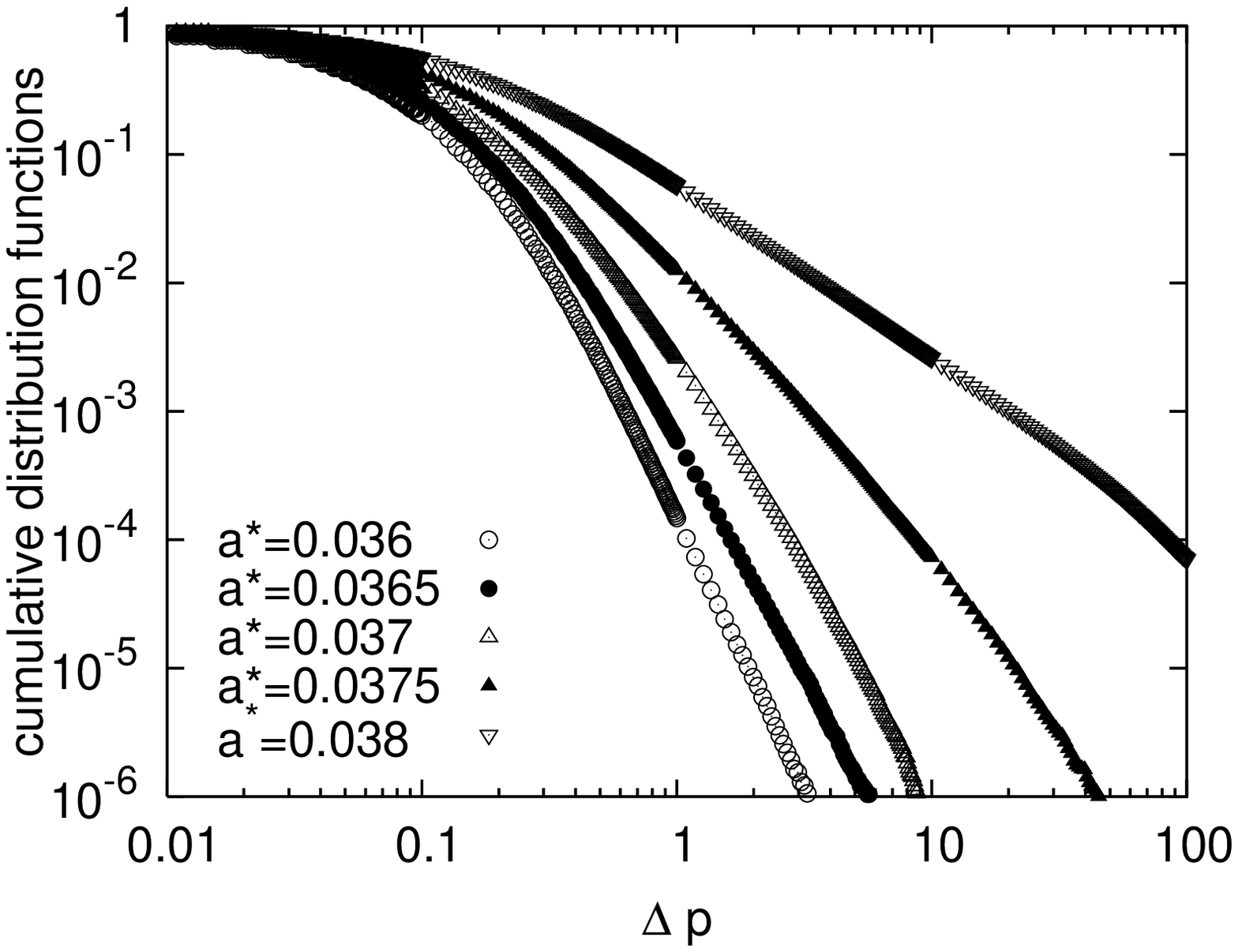}
\caption{Log-log plots of the cumulative distribution function of the
  price changes $\Delta p_s$. We fix $N=100$, $\Lambda=1.0$ and 
  $a^*=0.01$, and change $a^*$ (left). We fix $N=100$, $\Lambda=1.0$
  and $c^*=100.0$, and change $a^*$ (right). 
}
\label{fig:cdfs}
\end{figure}

\begin{figure}[htb]
\centering
\epsfxsize=320pt
\epsfbox{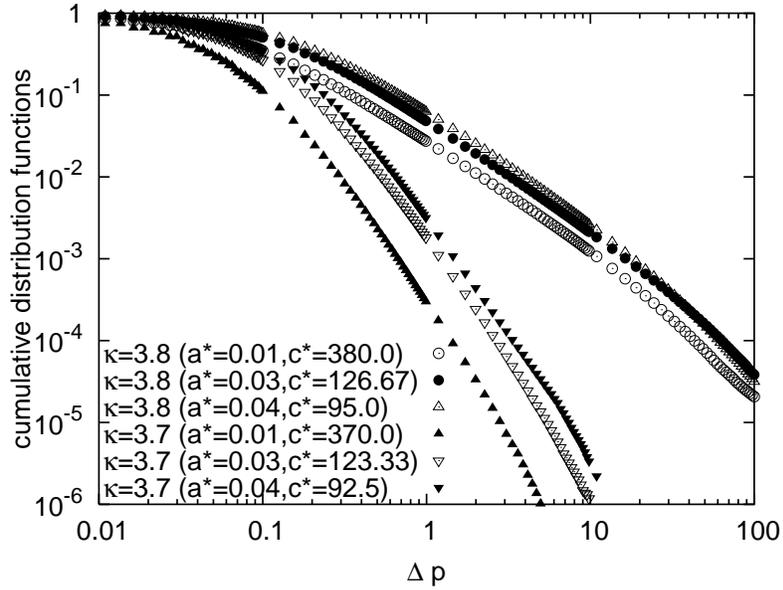}
\caption{Log-log plots of the cumulative distribution functions of the
  price changes $\Delta p_s$ at $\kappa = 3.8$ and $\kappa=3.7$. It
  shows the cumulative distribution functions at three different
  parameter sets for each $\kappa$.   
}
\label{fig:cdfs-same-kappa}
\end{figure}

Fig. \ref{fig:pdfs} displays semi-log plots of pdfs of $\Delta
p_s$ for various values of $a^*$ and $c^*$. The pdfs seem to have fat
tails and their tails depend on the values of both $a^*$ and
$c^*$. The corresponding cumulative distribution function (cdf) is
defined as, 
\begin{equation}
F(\geq|x|) = \int_{-\infty}^{-|x|}f(x')dx' +
\int_{|x|}^{\infty}f(x')dx',
\end{equation}
where $f(x)$ is a pdf. If the pdf follows power-law distribution the
corresponding cdf is given by  
\begin{equation}
F(\geq|x|)\propto|x|^{-\beta},
\end{equation}
where $\beta$ is a power law exponent ($\beta>0$), which is estimated
from a slope of the cdf in the double-log scale. As shown in fig. 
\ref{fig:cdfs} the power law exponent depends on the value of both
$c^*$ and $a^*$. For large $a^*$ and $c^*$ the power law exponent
becomes small. Moreover let be $\kappa \equiv a^*c^*$. Then the power law
exponent $\beta$ is a function of $\kappa$ as shown in
fig. \ref{fig:cdfs-same-kappa}. We discuss why the power law exponent
depends on $\kappa$ in the next section.

\begin{figure}[htb]
\centering
\epsfxsize=200pt
\epsfbox{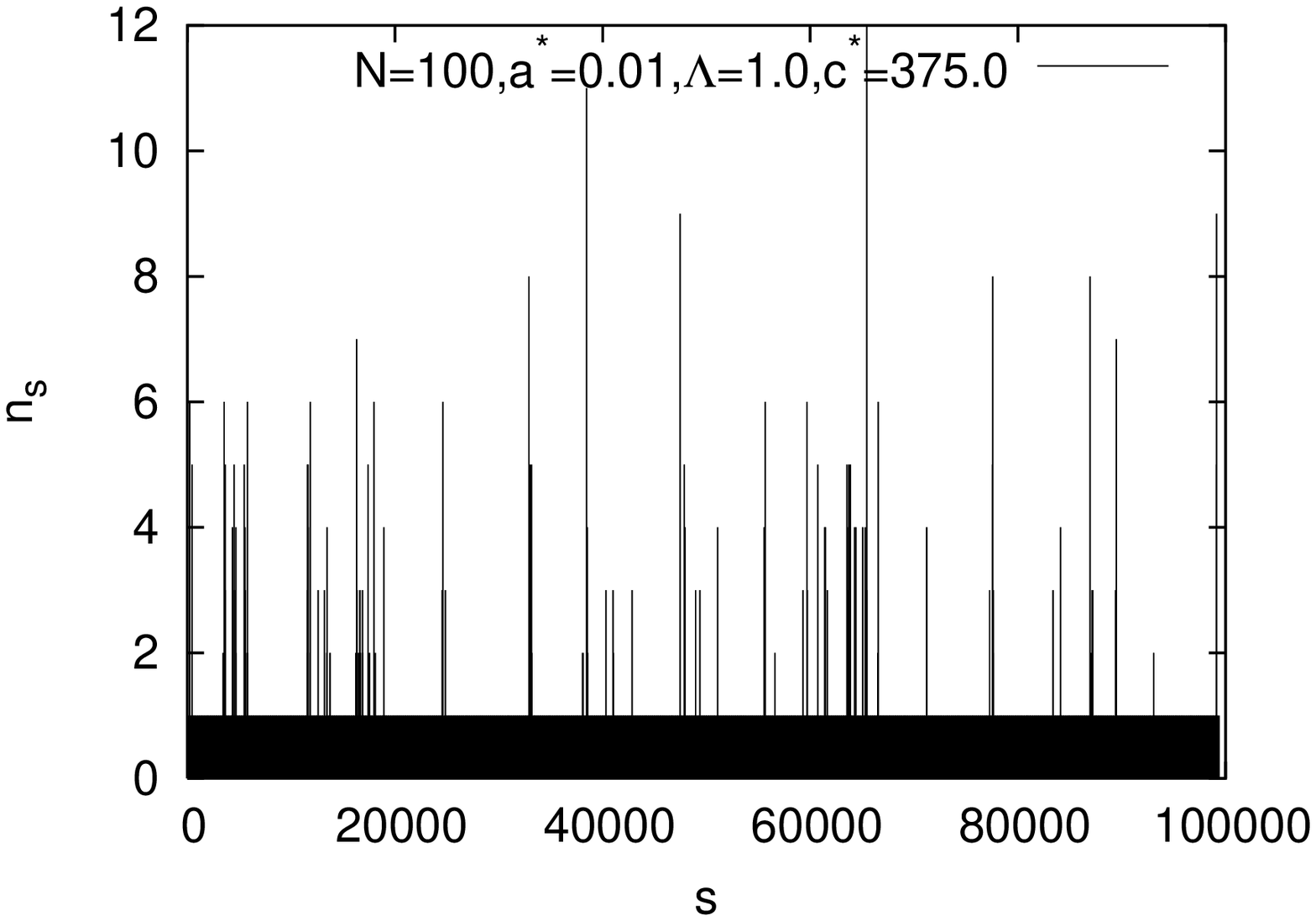}
\epsfxsize=200pt
\epsfbox{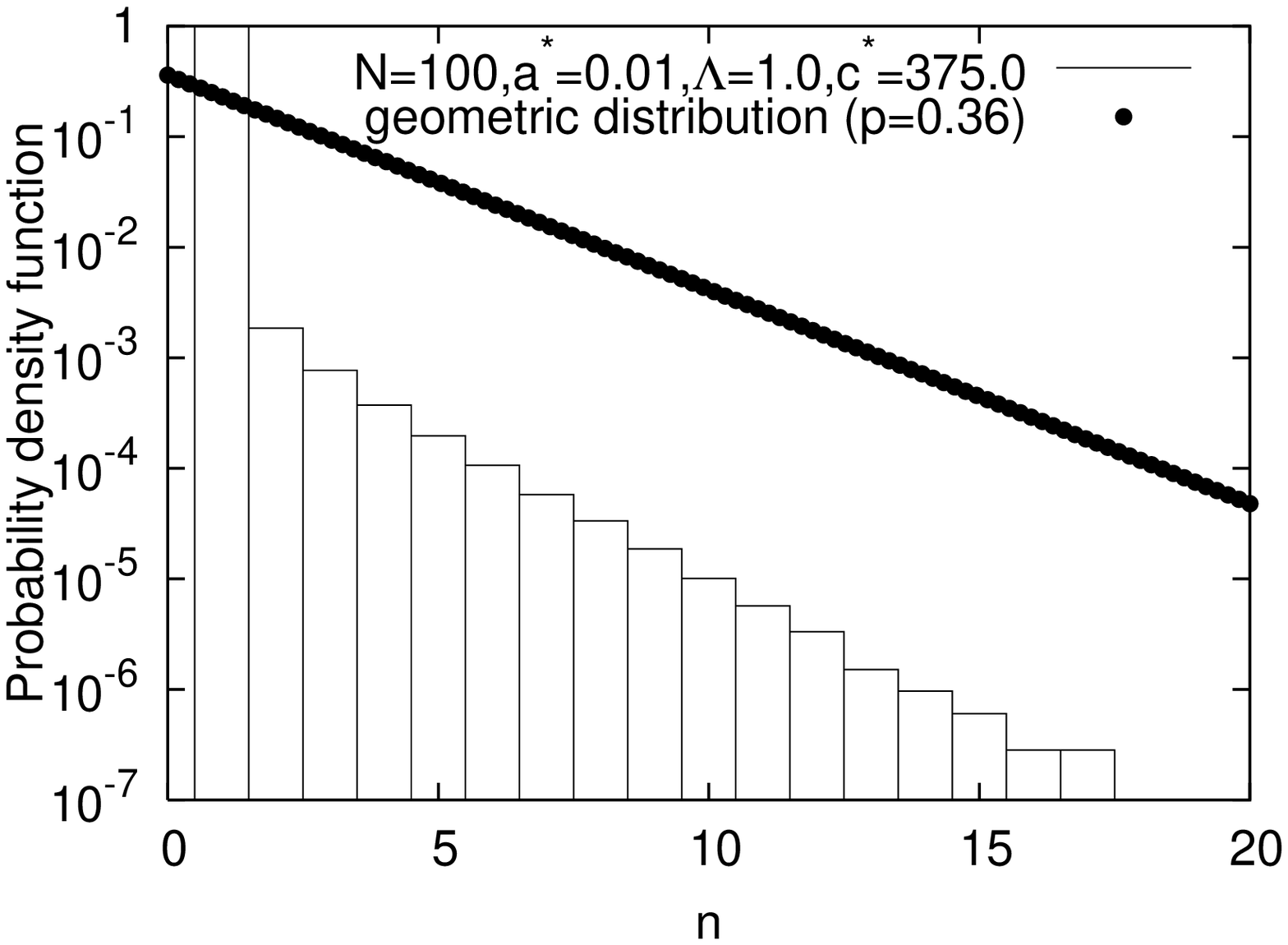}
\caption{A typical example of time series of an interval between
  successive trading $n_s$ (left) and its probability distribution
  function in semi-log scale (right). Parameters are fixed at $N=100$,
  $a^*=0.01$, $\Lambda=1.0$ and $c^*=375.0$.} 
\label{fig:n}
\end{figure}

Typical example of time series of $n_s$ and its pdf are shown in
fig. \ref{fig:n}, respectively. The pdf can be approximated by a
geometric distribution, 
\begin{equation}
W(n) = p(1-p)^n 
\label{eq:W}
\end{equation}
This means that occurrence of trading is fully random. By numerical
fitting in fig. \ref{fig:n} (right) we obtain $p=0.36$. From analysis
of high-frequency financial data it is clarified that the pdf
trading intervals at the same time is fitted by exponential
distribution~\cite{Takayasu:03}. 

\section{Stochastic approximation as the GARCH process}
\label{sec:ARCH-limit}
Here we consider why the pdf of the price changes has fat tails and that
its tail index depends on the parameter both $a^*$ and $c^*$. We will
show that the market price is approximately dominated by the GARCH
process through this section.  

\begin{figure}[htb]
\centering
\epsfxsize=320pt
\epsfbox{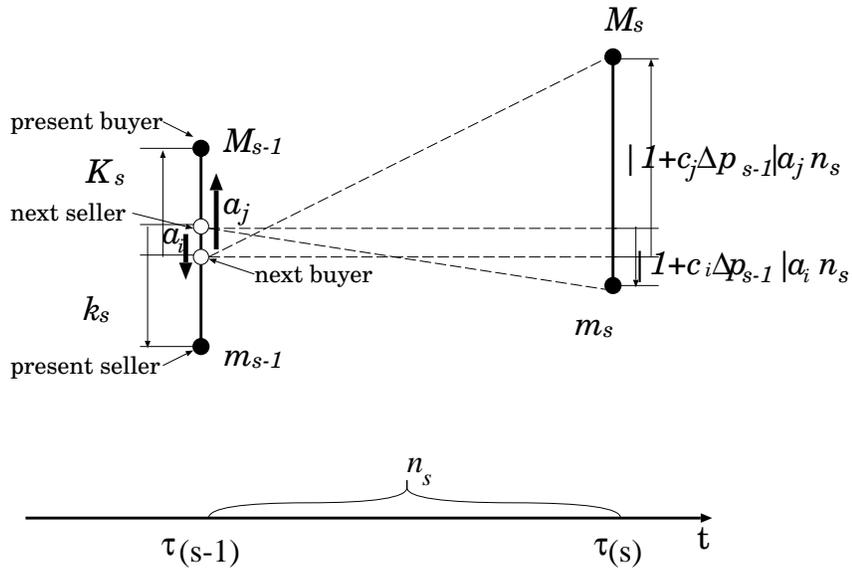}
\caption{A conceptual illustration for explaining dealers interaction between $\tau_{s-1}$ and $\tau_s$. $M_{s}$ represents a buy price at $\tau_s$, and $m_s$ a sell price. The Next buyer at $\tau_{s-1}$ adds $|1+c_j\Delta p_{s-1}|$ into his/her bid price $n_s$ times, and the next seller subtracts $|1+c_i\Delta p_{s-1}|$ from his/her bid price $n_s$ times.}
\label{fig:explanation-of-m_s-and-M_s}
\end{figure}

As shown by the numerical simulations the market price fluctuates
discontinuously. Here we consider $\Delta p_s$, a change of the market
price on the $s$th transaction as shown in fig. \ref{fig:fluctuation}. 
Let $M_s$ denote the buying price at the $s$th transaction, and $m_s$
the selling price as shown in fig. \ref{fig:explanation-of-m_s-and-M_s}. 
From the definition $P(\tau_s) = \frac{1}{2}(M_s + m_s)$ and $\Delta p_s =
P(\tau_s)-P(\tau_{s-1})$, we get 
\begin{eqnarray}
\nonumber
\Delta p_s &=& \frac{1}{2}(M_s + m_s) - \frac{1}{2}(M_{s-1}+m_{s-1}) 
\\
&=& \frac{1}{2}(M_s - M_{s-1}) + \frac{1}{2}(m_s - m_{s-1}).
\label{eq:Delta-p}
\end{eqnarray}
From eq. (\ref{eq:dealers-rule}) the next buyer at
$\tau_{s-1}$ adds $|1+c_j\Delta p_{s-1}|a_j$ into his/her bid price
$n_s$ times until he/she can trade, and the next seller
also subtracts $|1+c_i\Delta p_{s-1}|a_i$ from his/her ask price $n_s$
times. Hence the first term and the second term in eq. (\ref{eq:Delta-p})
are given by 
\begin{eqnarray}
M_s - M_{s-1} &=& |1+c_j\Delta p_{s-1}| a_{j} n_{s-1} - K_{s},
\label{eq:delta-M}
\\ 
m_s - m_{s-1} &=& k_{s}  - |1+c_i\Delta p_{s-1}| a_{i} n_{s-1},
\label{eq:delta-m}
\end{eqnarray}
where the subscript $j$ represents the dealer who gives the highest
bid price at the time $\tau_{s}$, and $i$ represents the dealer who
gives the lowest ask one. $K_{s}$ represents a difference of the bid
prices between the $(s-1)$th buyer and the $s$th buyer at
$\tau_{s-1}$, and $k_{s}$ of the ask prices between the 
$(s-1)$th seller and the $s$th seller.  

Substituting eqs. (\ref{eq:delta-m}) and (\ref{eq:delta-M}) into 
eq. (\ref{eq:Delta-p}) yields 
\begin{equation}
\Delta p_s = \frac{1}{2}|1+c_j\Delta p_{s-1}|a_j n_{s-1} - \frac{1}{2}
|1+c_i\Delta p_{s-1}|a_i n_{s-1} + \frac{1}{2}(k_s - K_s).
\label{eq:Delta-p-rewrite}
\end{equation}

\begin{figure}[htb]
\centering
\epsfxsize=200pt
\epsfbox{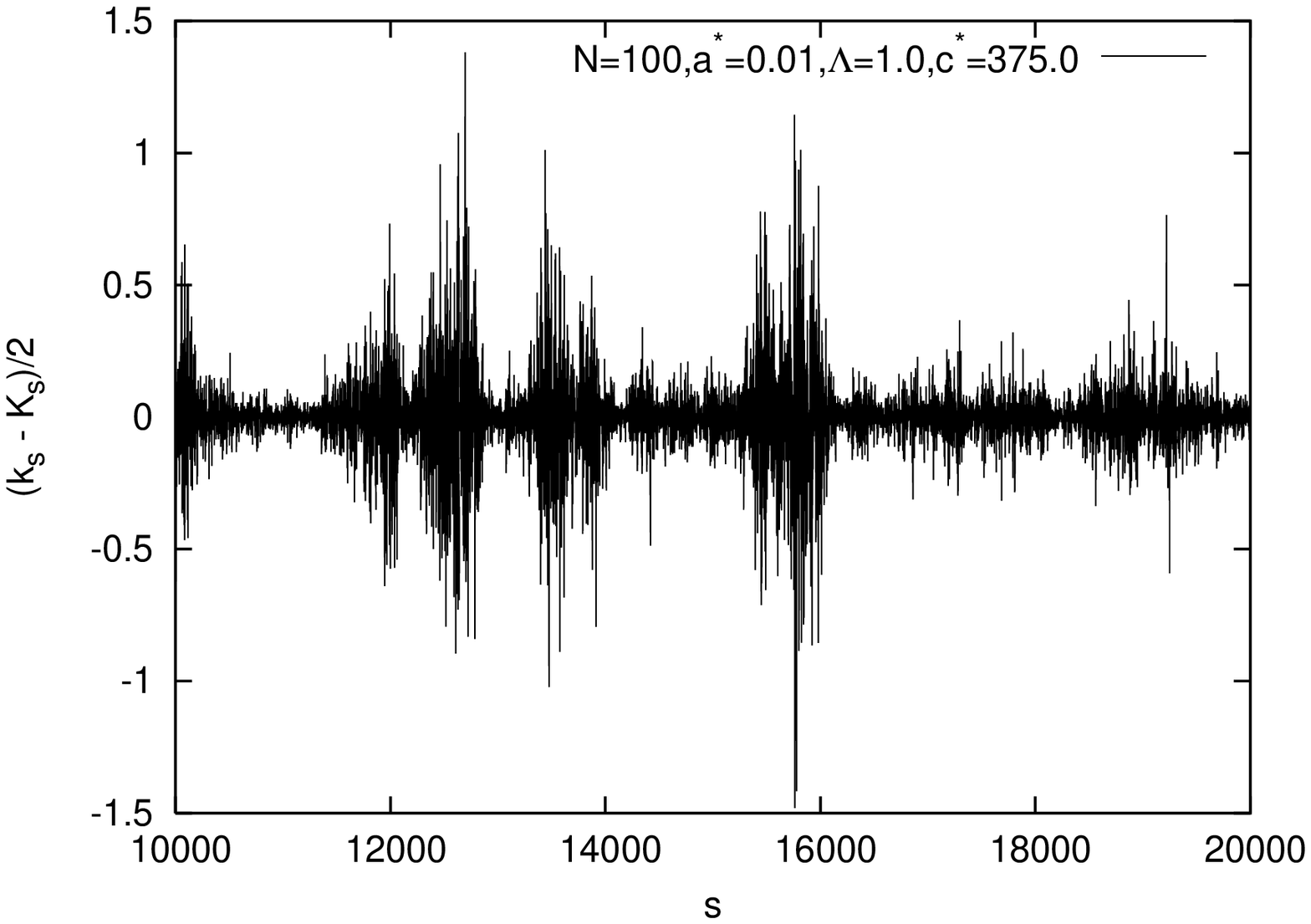}
\epsfxsize=200pt
\epsfbox{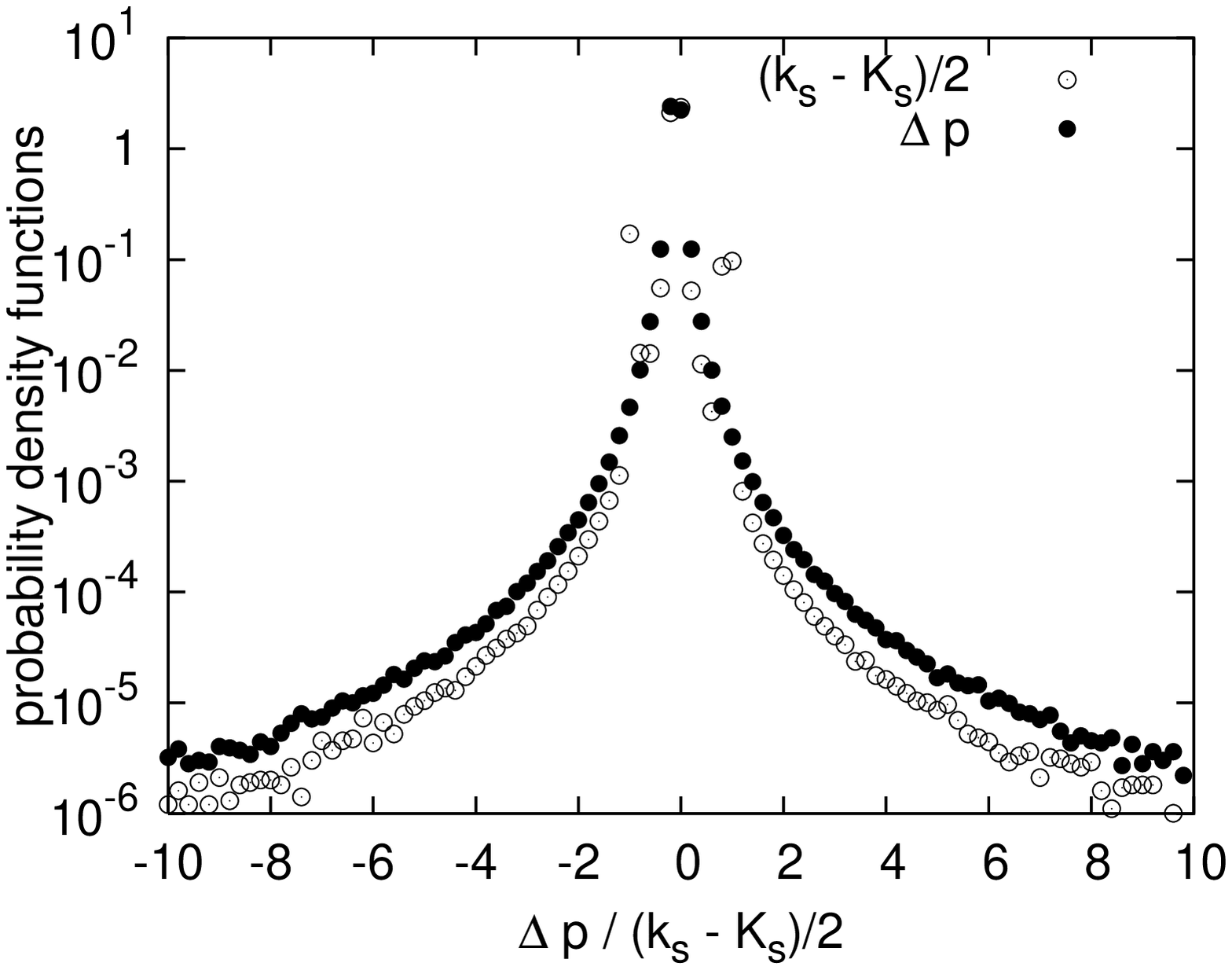}
\caption{A typical example of time series of $k_s - K_s$ at $N = 100$,
  $a^* = 0.01$, $\Lambda = 1.0$ and $c^* = 375.0$ (left). The pdf of $k_s - K_s$ and $\Delta p_s$ at the same parameters (right).}
\label{fig:additive-noise}
\end{figure}

Since the dealer number $N$ is large we assume that $\{c\}$ and
$\{a\}$ are mutually independent stochastic variable, which is uniformly
distributed in intervals $[-c^*,c^*]$ and $[0,a^*]$. Therefore we
have $\langle c \rangle = 0$, $\langle c^2 \rangle = c^{*2}/3$,
$\langle a \rangle = a^*/2$ and $\langle a^2 \rangle =
a^{*2}/3$. Moreover since the sellers and the buyers are symmetric
$k_s - K_s$ is also symmetric. Actually it is easily confirmed by 
a numerical simulation. Fig. \ref{fig:additive-noise} (left) is a
typical example of time series of $k_s - K_s$. It is obvious that $k_s
- K_s$ is symmetric. Therefore $\langle k_s - K_s \rangle = 0$.  

By taking conditionally averaging eq. (\ref{eq:Delta-p-rewrite}) over
dealers' indices $i$ and $j$ we have,
\begin{equation}
\langle \Delta p_{s} \rangle = 0.
\end{equation}

By taking square of eq. (\ref{eq:Delta-p-rewrite}) and averaging over
dealers' indices $i$ and $j$ under the condition that $\Delta
p_{s-1}$ is realized, we get  
\begin{eqnarray}
\nonumber
\lefteqn{\langle \Delta p_s ^2\rangle = \frac{1}{4}\langle (k_s -
  K_s)^2 \rangle + \frac{1}{4}\langle |1 + c\Delta p_{s-1}|^2 \rangle
  \langle a^2 \rangle \langle n_s^2 \rangle} \hspace{13cm} \\
\nonumber
\lefteqn{+\frac{1}{4}\langle |1 + c\Delta p_{s-1}|^2 \rangle \langle a^2
\rangle \langle n_s^2 \rangle - \frac{1}{2}\langle | 1 + c \Delta
p_{s-1}| \rangle \langle | 1 + c \Delta p_{s-1}| \rangle \langle a \rangle ^2
\langle n_s^2 \rangle} \hspace{13cm} \\
\lefteqn{=\frac{1}{4}\langle (k_s - K_s)^2 \rangle + \frac{1}{2} \langle |1
 + c\Delta p_{s-1}|^2 \rangle \langle a^2 \rangle \langle n_s^2
 \rangle - \frac{1}{2} \langle |1 + c \Delta p_{s-1}| \rangle^2
\langle a \rangle ^2 \langle n_s^2 \rangle}.\hspace{13cm}
\label{eq:Delta-p2}
\end{eqnarray}

The first term in the left hand side of eq. (\ref{eq:Delta-p2}) is not
constant. $k_s - K_s$ has large
fluctuations. Fig. \ref{fig:additive-noise} (right) is the pdf of $k_s 
- K_s$. The pdf is the same distribution as $\Delta p$. Hence we
assumed that $\langle (k_s - K_s)^2\rangle$ is proportional to a
conditional variance of $\Delta p_{s-1}$, namely 
\begin{equation}
\frac{1}{4} \langle (k_s - K_s)^2 \rangle = \eta \langle \Delta
p_{s-1}^2 \rangle, 
\end{equation}
where $\eta$ is a positive coefficient. The second term in the left
hand side of eq. (\ref{eq:Delta-p2}) is calculated as, 
\begin{equation}
\frac{1}{2} \langle |1 + c\Delta p_{s-1}|^2 \rangle \langle a^2 \rangle
\langle n_s^2 \rangle = \frac{a^{*2}}{6} \langle n_s^2 \rangle 
+ \frac{a^{*2}c^{*2}}{18} \Delta p_{s-1}^2 \langle n_s^2 \rangle. 
\end{equation}
The third term in the left hand side of eq. (\ref{eq:Delta-p2}) is
calculated as,  
\begin{equation}
\frac{1}{2} \langle | 1 + c \Delta p_{s-1} | \rangle^2 \langle a
\rangle^2 \langle n_s^2 \rangle = 
\left\{
\begin{array}{ll}
\frac{a^{*2}}{32}(\frac{1}{c^*\Delta p_{s-1}} + c^*\Delta p_{s-1})^2
\langle n_s^2 \rangle & (|\Delta p_{s-1}| \geq 1/c^*) \\
\frac{a^{*2}}{32} \langle n_s^2 \rangle  & (|\Delta p_{s-1}| < 1/c^*) 
\end{array}
\right..
\end{equation}

Therefore by using $\sigma_s^2 = \langle \Delta p_s^2 \rangle - \langle
\Delta p_s \rangle^2$ (\ref{eq:Delta-p2}) is described as, 
\begin{equation}
\sigma_s^2 = 
\left\{
\begin{array}{ll}
\omega_1 + \eta \sigma_{s-1}^2 + \rho_1 \Delta p_{s-1}^2
+ \gamma \frac{1}{\Delta p_{s-1}^2} &  (|\Delta p_{s-1}| \geq 1/c^*) \\
\omega_2 + \eta \sigma_{s-1}^2 + \rho_2 \Delta p_{s-1}^2
& (|\Delta p_{s-1}| < 1/c^*) 
\end{array}
\right.,
\label{eq:GARCH-limit}
\end{equation} 
where $\omega_1 \equiv \frac{5}{48} a^{*2} \langle n_s^2 \rangle$,
$\rho_1 \equiv \frac{7}{288}\kappa^2 \langle n_s^2 \rangle$,
$\gamma \equiv \frac{a^{*2}}{32 c^{*2}} \langle n_s^2 \rangle$,
$\omega_2 = \frac{1}{24} a^{*2} \langle n_s^2 \rangle$ and $\rho_2 
\equiv \frac{1}{18} \kappa^2 \langle n_s^2 \rangle$. From
eq. (\ref{eq:W}) we obtain $\langle n_s^2 \rangle = \frac{(2-p)(1-p)}{p^2}
\approx 2.92$

For large $|\Delta p_{s-1}|$ the term $\gamma \frac{1}{\Delta
  p_{s-1}^2}$ of eq. (\ref{eq:GARCH-limit}) is almost zero. Then
eq. (\ref{eq:GARCH-limit}) is approximated as a GARCH(1,1) process. 
The probability density function of the GARCH(1,1) process has fat
tail~\cite{Ghose:95,Groenendijik:95}. The power law exponent is a
function of $\sigma_1$ and $\eta$. Namely the power law exponent depends on
$\kappa$. Furthermore if $\kappa$ is small then the
second term of eq. (\ref{eq:GARCH-limit}) vanishes, and we obtain
$\sigma_s^2 = \omega_2 + \eta \sigma_{s-1} ^2 $. The result of the
  iteration is given by, 
\begin{equation}
\sigma_s^2 \rightarrow \frac{\sigma_0^2}{1-\eta} + \frac{\omega_2
  \eta}{(1-\eta)^2} \mbox{\hspace{3ex}($s\rightarrow\infty$)}.
\end{equation}
Namely the variance of $\Delta p_s$ is nearly constant. Then $p_s$
behaves similarly to a Brownian motion. 

\section{Discussion}
\label{sec:discussion}
\subsection{Correlation function}
\begin{figure}[htb]
\centering
\epsfxsize=200pt
\epsfbox{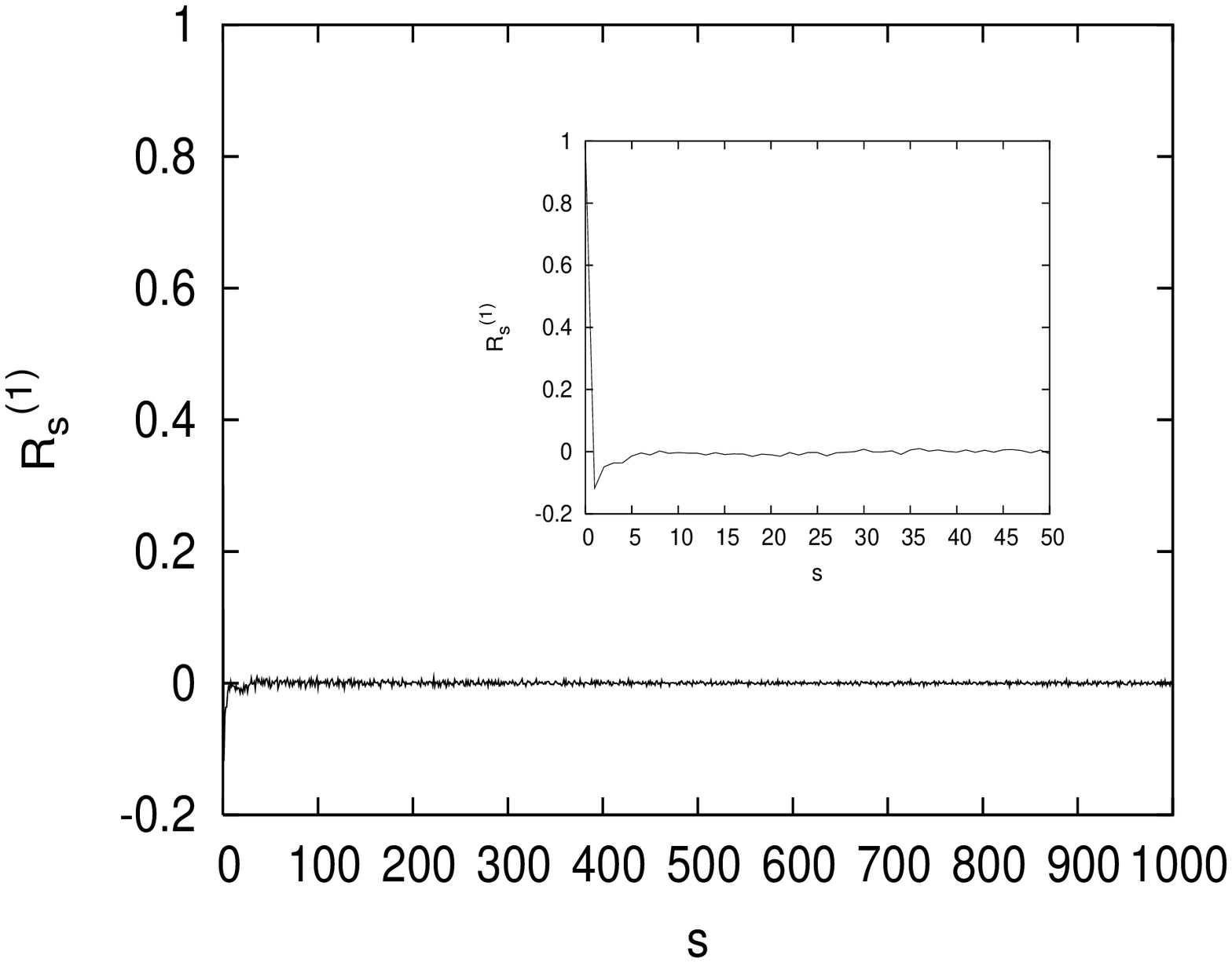}
\epsfxsize=200pt
\epsfbox{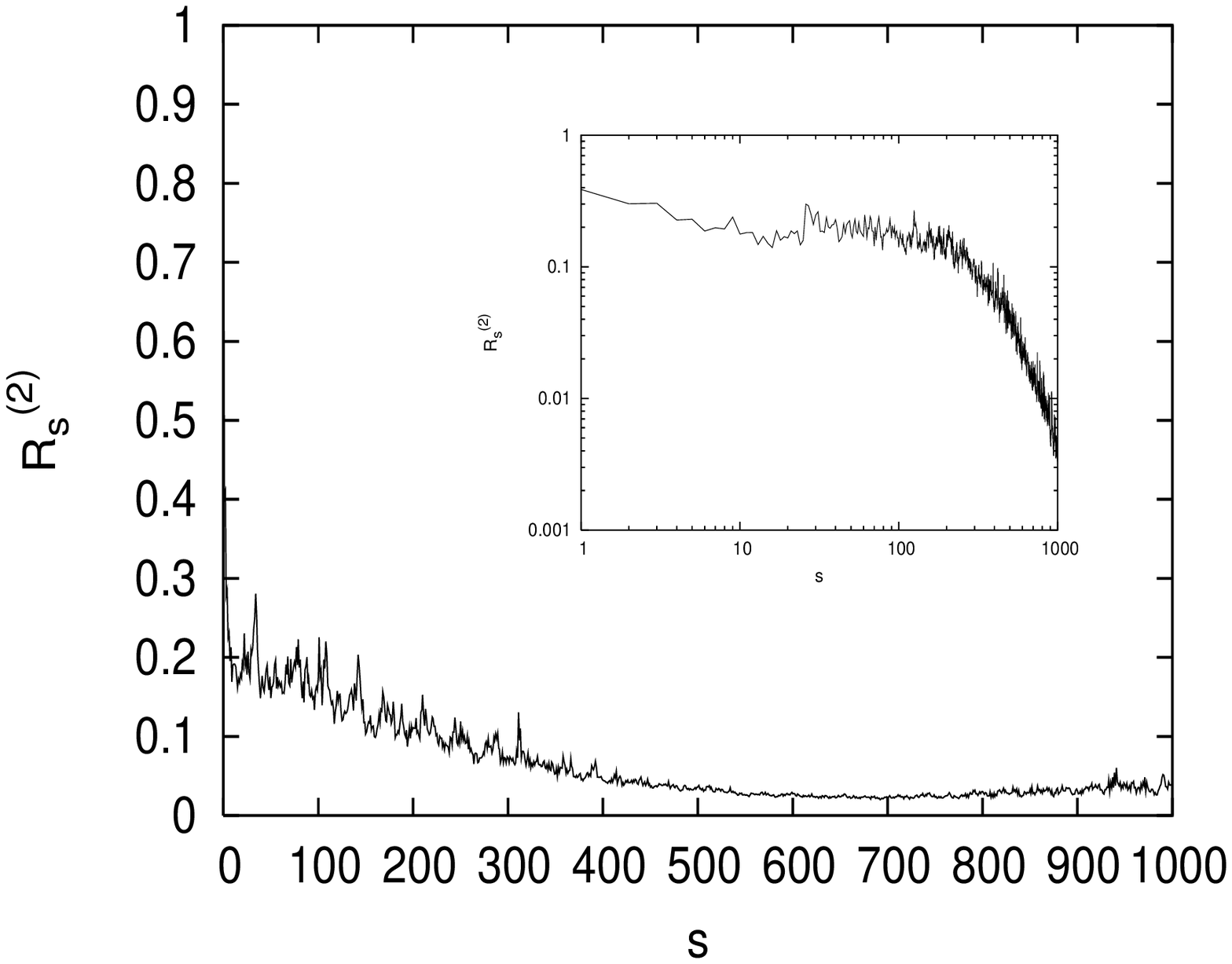}
\caption{Autocorrelation coefficient of price changes
  (left) and Autocorrelation coefficient of squared price 
  changes (right) are calculated from temporal development of price
  changes. Parameters are fixed at $N=100$, $a^*=0.01$,
  $\Lambda=1.0$ and $c^*=375.0$.}
\label{fig:correlation}
\end{figure}

We calculate an autocorrelation coefficient of the market price changes
and one of squared changes. The volatility clustering is the
well-known fact that the squared changes of the market price are
clustering in financial data. It is indicated that the volatility
clustering is related to a long time correlation of the
market~\cite{Pasquini:99}. These two autocorrelation coefficients are
defined by
\begin{eqnarray}
R^{(1)}_s &=& \frac{\langle \Delta p_{t+s} \Delta p_{t} \rangle -
  \langle \Delta p_{t+s} \rangle \langle \Delta p_{t} \rangle}{\langle
  \Delta p_{t}^2 \rangle - \langle \Delta p_{t} \rangle^2},
  \\
R^{(2)}_s &=& \frac{\langle \Delta p_{t+s}^2 \Delta p_{t}^2 \rangle -
  \langle \Delta p_{t+s}^2 \rangle \langle \Delta p_{t}^2 \rangle}
{\langle \Delta p_{t}^4 \rangle - \langle \Delta p_{t}^2 \rangle^2}.
\end{eqnarray}
As shown in fig. \ref{fig:correlation} (left) the autocorrelation
coefficient is negative at small $s$. It means that the
price change tends to move to an opposite direction of the last price
change. On the other hand autocorrelation coefficient of the squared
change in fig. \ref{fig:correlation} (right) has a long time
correlation. It is interesting that the volatility of price changes
has long time correlation although each dealer just depends on the latest
price change. Actually it is reported that the volatility ocorrelation
function of the GARCH(1,1) process decreases exponentially~\cite{Ding:96}. 
However that is only true when a ARCH(1) parameter plus a GARCH(1)
parameter is less than or equal to 1. When the ARCH(1) parameter plus
the GARCH(1) parameter is greater than 1 it is impossible to derive
the autocorrelation function in the same analytical way. From
numerical simulations of the pure GARCH(1,1) process we confirm that
the autocorrelation function has long tail when the ARCH(1) parameter
plus the GARCH(1) parameter is greater than 1. Specifically its long
tail is significant when the GARCH(1) parameter is near 1. 
Hence we can explain from this property that the market price
fluctuations of the dealer model exhibits the volatility clustering
when $\eta + \rho_1 > 1$ and $\eta$ is near 1.

\subsection{Demand and Supply curve of the model}
\begin{figure}[htb]
\centering
\epsfxsize=320pt
\epsfbox{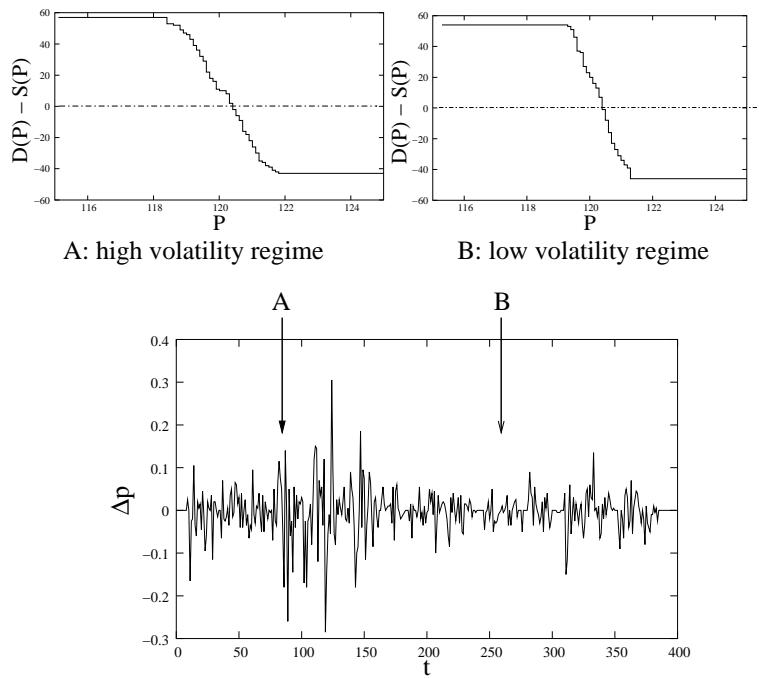}
\caption{A difference between cumulative frequency distribution
  $D(P,t)$ and $S(P,t)$. Capital A exhibits a high volatility regime,
  and B a low volatility one. Parameters are fixed at $N=100$,
  $a^*=0.01$, $\Lambda=1.0$ and $c^*=375.0$.}
\label{fig:dscurve}
\end{figure}

In the model introduced in the article we treat the case that demand and
supply are automatically balancing, so that the total number of both sellers
and buyers is conserved. We introduce time-dependent density of
buyers and sellers for a market price $P$ at time $t$, $d(P,t)$ and
$s(P,t)$, respectively~\cite{Takayasu:99}. The cumulative frequency
distributions of sellers and buyers $D(P,t)$ and $S(P,t)$ are
respectively defined by
\begin{eqnarray}
D(P,t) &=& \int_{0}^{P}d(P',t)dP' \\
S(P,t) &=& \int_{P}^{\infty}s(P',t)dP'.
\end{eqnarray}
Of course in practical markets numbers of sellers and buyers can be
observed partially. However in the numerical simulation one can
calculate them from all the dealers' variable. Here we consider
relation between volatility and a demand-supply
curve. Fig. \ref{fig:dscurve} shows $D(P,t) - S(P,t)$ in high
volatility regime (A) and in low volatility one (B). It is found that
a slope of $D(P,t) - S(P,t)$ in the high volatility regime is
gradual. On the other hand one in the low is rapid. This means that a
slope of $D(P,t) - S(P,t)$ is related to the volatility. Hence the
volatility clustering is attributed to gradually changing of the
demand-supply curve.

\subsection{Amount of sellers and amount of buyers}
In the model the seller and buyer who have exchanged commodity open
his/her position depending on the latest price change after they make
a trade. We assume that the sellers go on selling when the market price goes
down and that the buyers go on buying when the market price goes up.
Let be $N_S$ and $N_B$ represent an amount of the sellers and one of buyers.
Then $N_S - N_B$ fluctuates around zero. The reason is as follows:
When $N_S - N_B > 0$ the market price tends to go down. When $N_S -
N_B < 0$ it goes up. However when the market goes down $N_S$ decreases
and $N_B$ increases. $N_S - N_B$ becomes less than zero. When the
market price goes up $N_S$ increases and $N_B$ decreases. $N_S - N_B$
becomes greater than zero. If the sellers/buyers go on selling/buying
when the market price goes up/down then a crash/bubble
occurs. Throughout both the GARCH process and the
demand and supply we think that there are two reasons why the market
price fluctuates. One results from rules of a dealers'
prediction depending on the past market price. Another results from
unbalancing of demand and supply. 

\section{Conclusion}
\label{sec:conclusion}
We introduced the dealer model in which a commodity is exchanged
such as a foreign exchange market. The model generates the time series
statistically similar to real financial one although the model has a
few parameters. This means that the model contains primitive factor
for the market price fluctuations. The probability density
function of the artificial price changes has power law tails. The
autocorrelation function of the changes has anti-correlation for a few
ticks. This implies that the market price changes tend to move in
opposite. A squared autocorrelation function (volatility correlation
function) has a long tail although the dealers depend on the latest
price change. We derived the GARCH process for price changes under the
assumption that the dealer homogeneously trades irrespective of their
coefficients {\it i.e.} we can take average of $\Delta p_s^2$ over all
the dealers' indices. From the GARCH(1,1) limit we explained the
power-law of the probability density function and the long time
correlation of volatility correlation function. The long time
correlation of volatility can be also seen in a demand-supply
curve. The slope of the demand-supply curve is gradual in high
volatility regime. In contrary the slope of demand-supply curve is
rapid in a low volatility case. We guess that the market price
fluctuations result from both the dealers' prediction depending on the
past market prices and unbalancing of demand and supply. 

The relation between microscopic behavior of dealers and macroscopic
feature of a market will be bridged by statistical mechanics. Then the
intriguing problem is to clarify what kind of microscopic interaction
between dealers generate statistical feature of macroscopic variables.

\section*{Acknowledgements}
One of the authors (A.-H. Sato) thanks T. Munakata and M. Aoki for
stimulative discussions and useful comments.

\begin{table}[h]
\caption{Model parameters; $N$, $a^*$, $\Lambda$ and $c^*$.}
\label{tab:parameters}
\centering{
\begin{tabular}{|l|l|}
\hline
parameter & detail \\ \hline 
$N$ & amount of the dealers \\ \hline 
$a^*$ & a prediction coefficient $a_i$ \\ \hline 
$\Lambda$ & difference between retried ask/bid price and the market price. 
\\ \hline 
$c^*$ & for a prediction coefficient $c_i$ \\
\hline
\end{tabular}
}
\end{table}
\end{document}